\documentclass[sigconf]{acmart}

\renewcommand\footnotetextcopyrightpermission[1]{} 

\AtBeginDocument{%
  \providecommand\BibTeX{{%
    \normalfont B\kern-0.5em{\scshape i\kern-0.25em b}\kern-0.8em\TeX}}}

\AtBeginDocument{%
 \providecommand\BibTeX{{%
    \normalfont B\kern-0.5em{\scshape i\kern-0.25em b}\kern-0.8em\TeX}}}
    
\usepackage{enumitem}
\usepackage{multirow}
\usepackage{makecell}
\usepackage{amsthm}
\usepackage{amsfonts}
\usepackage{url}

\newcommand{\ie}{\emph{i.e., }}

\newcommand{\wrt}{\emph{w.r.t. }}

\newcommand{\aka}{\emph{aka. }}

\begin{document}

\title{SCTc-TE: A Comprehensive Formulation and Benchmark for Temporal Event Forecasting}

\author{Yunshan Ma}
\authornote{Equal contricution.}
\affiliation{
     \institution{National University of Singapore}
     \country{}
}
\email{yunshan.ma@u.nus.edu}

\author{Chenchen Ye}
\authornotemark[1]
\affiliation{
     \institution{University of California, Los Angeles}
     \country{}
}
\email{ccye@cs.ucla.edu}

\author{Zijian Wu}
\authornotemark[1]
\affiliation{
     \institution{National University of Singapore}
     \country{}
}
\email{zijian.wu@u.nus.edu}

\author{Xiang Wang}
\authornote{Corresponding author. Xiang Wang is also affiliated with Institute of Artificial Intelligence, Institute of Dataspace, Hefei Comprehensive National Science Center.}
\affiliation{
     \institution{University of Science and Technology of China}
     \country{}
}
\email{xiangwang1223@gmail.com}

\author{Yixin Cao}
\affiliation{
     \institution{Singapore Management University}
     \country{}
}
\email{caoyixin2011@gmail.com}

\author{Liang Pang}
\affiliation{
     \institution{Institute of Computing Technology, Chinese Academy of Sciences}
     \country{}
}
\email{pangliang@ict.ac.cn}

\author{Tat-Seng Chua}
\affiliation{
     \institution{National University of Singapore}
     \country{}
}
\email{dcscts@nus.edu.sg}

\begin{abstract}
Temporal complex event forecasting aims to predict the future events given the observed events from history. Most formulations of temporal complex event are unstructured or without extensive temporal information, resulting in inferior representations and limited forecasting capabilities. To bridge these gaps, we innovatively introduce the formulation of Structured, Complex, and Time-complete temporal event (SCTc-TE). Following this comprehensive formulation, we develop a fully automated pipeline and construct a large-scale dataset named MidEast-TE from about 0.6 million news articles. This dataset focuses on the cooperation and conflict events among countries mainly in the MidEast region from 2015 to 2022. Not limited to the dataset construction, more importantly, we advance the forecasting methods by discriminating the crucial roles of various contextual information, \ie local and global contexts. Thereby, we propose a novel method LoGo that is able to take advantage of both Local and Global contexts for SCTc-TE forecasting. We evaluate our proposed approach on both our proposed MidEast-TE dataset and the original GDELT-TE dataset. Experimental results demonstrate the effectiveness of our forecasting model LoGo. The code and datasets are released via \url{https://github.com/yecchen/GDELT-ComplexEvent}.
\end{abstract}



\maketitle
\pagestyle{plain}

\section{Introduction}

\textbf{T}emporal \textbf{E}vent (TE) forecasting conceptually targets at predicting a future event based on observed facts from history. People seek to mine the rules that govern the evolution of various events, in order to facilitate disaster prevention or early warning in various areas, such as civil unrests or regional conflicts. Due to its significant value, event forecasting has garnered more and more interests from the research communities.

\begin{figure}
    \centering
    \includegraphics[width = 0.95\linewidth]{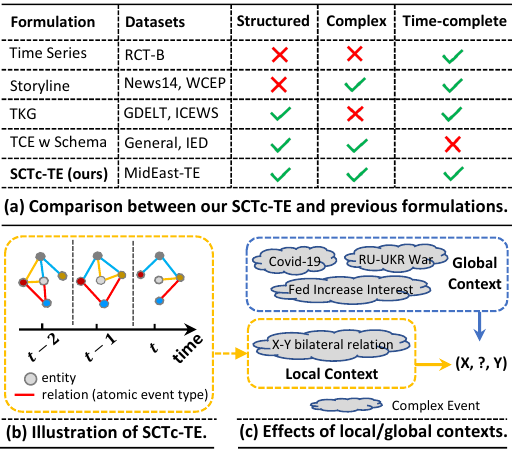}
    \vspace{-0.1in}
    \caption{(a) Comparison between our SCTc-TE and previous TE formulations; (b) An illustration of our SCTc-TE formulation; (c) The motivation of leveraging both local and global contexts.}
    \label{fig:motivation}
    \vspace{-0.15in}
\end{figure}

Albeit the previous diverse formulations of TE forecasting, we highlight three key properties that are crucial for both representation and forecasting of TE, \ie \textbf{Structured}, \textbf{Complex}, and \textbf{Time-complete}. First, structured representations, such as Temporal Knowledge Graph (TKG)~\cite{TKG-survey} or Temporal Complex Event (TCE) with schema~\cite{theFuture}, are simple and flexible for indexing and query, and the concise format of structured data enables the storing of large-scale events with limited memory. In contrast, other unstructured formulations, such as time series~\cite{RCT_B} or natural language~\cite{eventStoryLine}, are either non-flexible to represent multi-line of events or redundant due to natural language representation. Second, Complex Event (CE), which is composed by a set of atomic events~\cite{theFuture}, is capable of capturing multiple actors, relations, and timelines, thus satisfying various requirements in practice. Conversely, the formulations of atomic event, such as GDELT~\cite{gdelt} or ACE2005~\cite{ACE2005}, are restricted to individual events that cannot model the complex scenarios or even perform forecasting. Third and more importantly,  the time-complete characteristic, which means that every atomic event in a certain complex event has its corresponding timestamp, is indispensible. Some previous works~\cite{theFuture} define a specific \textit{temporal relation} to retain the temporal information within a complex event; however, it requires $O(N^2)$ temporal relations in order to fully preserve the temporal information of a complex event that has $N$ atomic events. As a result, the temporal relations are rarely complete due to the quadratic relation space. In the summary as shown in Figure~\ref{fig:motivation} (a), to the best of our knowledge, none of previous formulations satisfy all the three properties.     


\begin{figure*}[t]
    \centering
    \includegraphics[width = 0.96\linewidth]{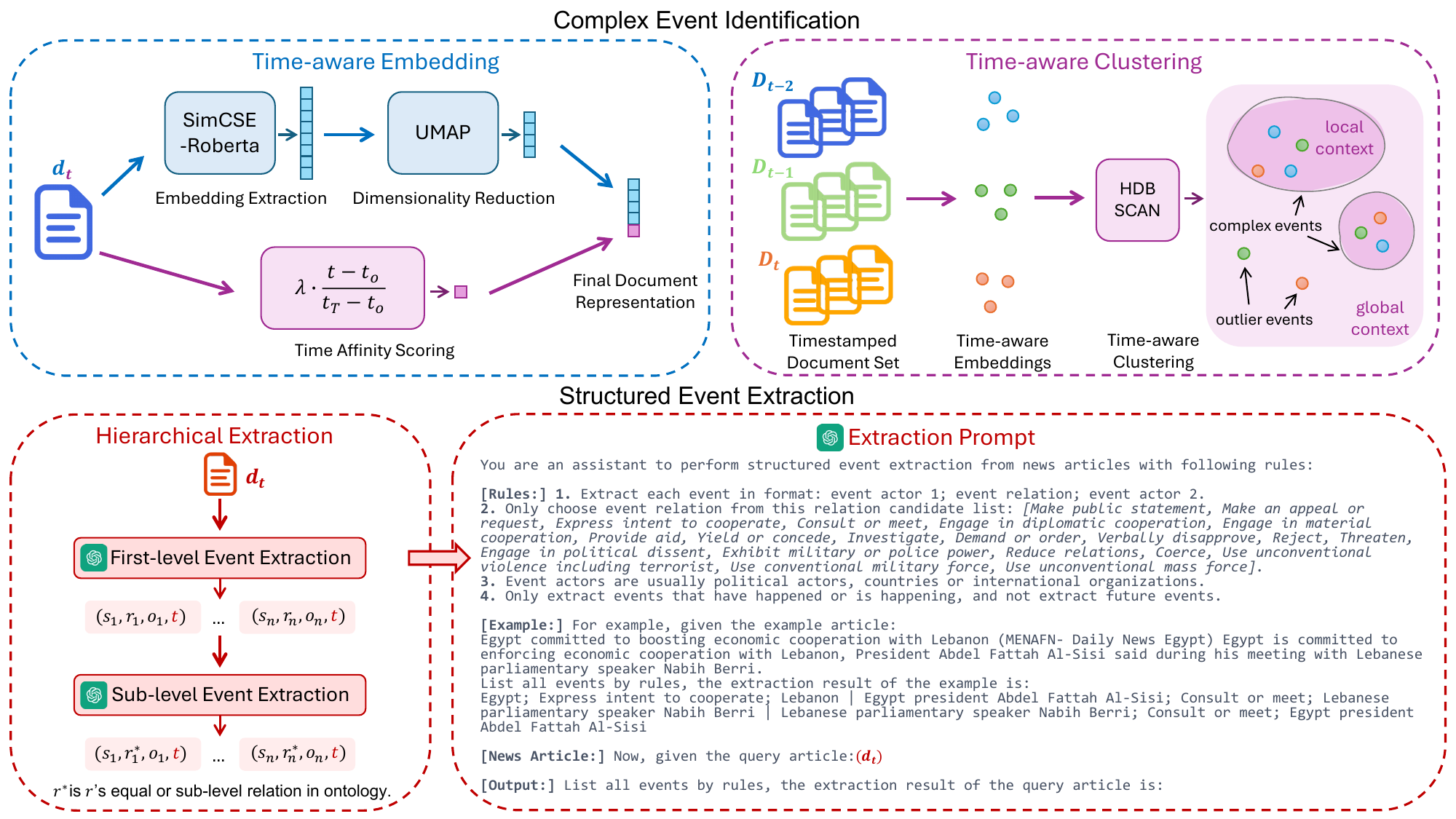}
    \caption{The dataset construction pipeline of SCTs-TE consists of: 1) \textit{Complex Event Identification} clusters complex events by time-aware document embedding and 2) \textit{Structured Event Extraction} hierarchically extracts structured events with LLMs.}
    \label{fig:pipeline}
\end{figure*}

To bridge the gap, we borrow the merits of both TKG and TCE with schema and propose a novel formulation to represent the \textbf{S}tructured, \textbf{C}omplex, and \textbf{T}ime-\textbf{c}omplete \textbf{T}emproal \textbf{E}vent (SCTc-TE). Specifically, as shown in Figure~\ref{fig:motivation} (b), we define one SCTc-TE as a list of semantically related and chronologically ordered graphs, each of which consists of a set of atomic events that are occurring at the same timestamp. We extend the atomic event formulation from the quadruple in GDELT~\cite{gdelt} to a quintuple $(s, r, o, t, c)$, where $s$, $r$, $o$, $t$ and $c$ represents the subject, relation~\footnote{Here relation refers to the atomic event type.}, object, timestamp, and CE, respectively. In contrast to TKG, where all the atomic events are scattered, our formulation groups them into clusters, and each cluster corresponds to a CE. More importantly, each atomic event has an absolute timestamp, endowing our formulation with the time-complete property. In order to implement this new formulation, we develop a simple and fully automated pipeline, which utilizes the pre-trained large language models (LLMs) and time-aware clustering, to construct SCTc-TEs from news articles. Based on this pipeline, we construct two large-scale event forecasting datasets, MidEast-TE and GDELT-TE, from 0.6 million news articles. 

Based on the newly proposed formulation and constructed datasets, we design a novel TE forecasting method that is able to leverage both local and global contexts. Given a particular query $(s, r, ?, t, c)$ to be forecast, we term the historical events belonging to the same complex event as the local context and all the historical events as the global context. The local context preserves the most relevant information that can be used to perform forecasting; while, the global context, which is usually noisy and scattered, provides a universal background that can also affect the evolution of certain complex events. For the example shown in Figure~\ref{fig:motivation} (c), given the complex event that mainly depicts the bilateral relationship between countries X and Y, both the historical relations between these two countries and the global contexts, such as Covid-19 pandemic, RU-UKR war, and Federate Reserve increase interest, will affect the bilateral relations. However, previous approaches solely rely on either the global context~\cite{RENET,REGCN} or the local context~\cite{theFuture}, which is sub-optimal in forecasting complex events. In this work, we propose to unify the modeling of both local and global contexts for TE forecasting. Concretely, we adopt two context learning modules to separately learn the entity and relation representations under the local and global context, respectively. An early-fusion strategy followed by a decoder is then applied to achieve the final forecasting. Extensive experiments demonstrate that our model outperforms SOTA methods. The main contributions of this work are as follows:

\begin{itemize}[leftmargin=*]
    \item To the best of our knowledge, we are the first to propose the SCTc-TE formulation that encompasses all the structured, complex, and time-complete properties of TEs.
    \item Based on the SCT-TCE formulation, we develop a fully automated pipeline to construct TEs from news articles and construct two large-scale datasets, MidEast-TE and GDELT-TE. 
    \item We propose a novel method (LoGo) that captures both the local and global contexts for SCTc-TE forecasting, and extensive evaluations and experiments demonstrate the richness of our datasets and the effectiveness of the proposed method. 
\end{itemize}
\section{Problem Formulation and Dataset} \label{sec:preliminary}

We first formally define the SCTc-TE and its forecasting task. Second, we develop a pipeline to construct SCTs-TE from news articles, as shown in Figure~\ref{fig:pipeline}. 

\subsection{Problem Formulation} \label{sebsec:prob_form}

We define one SCTc-TE as a list of timestamped graphs $\mathbf{G}^c=[G_1^c, G_2^c, \cdots, G_t^c]$, where $c \in \mathcal{C}$ is the identifier of one specific \textbf{Complex Event} (CE) and $\mathcal{C}$ is the entire identifier set for all CEs; and each graph is defined as $G_t^c=\{(s_n, r_n, o_n, t, c)\}_{n=1}^{N_t^c}$, where $(s_n, r_n, o_n, t, c)$ is the $n$-th \textbf{atomic event} in $G_t^c$ and $N_t^c$ is the number of atomic events at timestamp $t$ for the CE $c$. In terms of each atomic event, $s \in \mathcal{E}$, $r \in \mathcal{R}$, and $o \in \mathcal{E}$ correspond to the subject entity, relation, and object entity, respectively; $t$ is the timestamp when the certain atomic event occurs; $\mathcal{E}$ and $\mathcal{R}$ are the entity and relation set, respectively. Typically, $\mathbf{G}^c \in \mathcal{G}$ is constructed based on a list of timestamped document set $\mathbf{D}^c=[D_1^c, D_2^c, \cdots, D_t^c] \in \mathcal{D}$, where $\mathbf{G}^c$ is the \textbf{local context} for all the atomic events that the CE $c$ contains, while $\mathcal{G}$ is the \textbf{global context} that is a combined graph of all the CEs $\mathcal{C}$. $\mathcal{D}$ is the entire document set that consists of a list of timestamped document set $[\mathcal{D}_1, \mathcal{D}_1, \cdots, \mathcal{D}_t]$. Note that, both $D_t^c$ and $\mathcal{D}_t$ are a set of documents. 

Given a set of document $\mathcal{D}$, the \textbf{SCTc-TE Construction} task aims to identify the CEs, \ie $\mathbf{D}^c$, from which we then extract the atomic events and form the SCTc-TE graph $\mathbf{G}^c$. Given a partial CE $\mathbf{G}_{\le t}^c=[G_1^c, G_2^c, \cdots, G_t^c]$ and one query $(s, r, ?, t+1, c)$ at the next timestamp, the \textbf{SCTc-TE Forecasting} aims to predict the object $o$.

\subsection{SCTc-TE Construction Pipeline} \label{subsec:tce_construction}

We first introduce the domain and data sources that we used to illustrate the pipeline. Following on, we explicate how to perform event extraction using LLMs, then we present how to identify CEs from news articles, and finally show the statistics and evaluations of our datasets.

\subsubsection{Domain and Data Sources} We construct our datasets based on the GDELT~\cite{gdelt} corpus, which is a large-scale TKG dataset that has publicly accessible URLs of news articles. More importantly, it follows a well-defined ontology, \ie CAMEO~\cite{CAMEO}, which is authentic in describing international political events and curated by well-known domain experts. As the original GDELT dataset is very large, we follow the previous approaches~\cite{Glean,CMF} and crop a subset that is from three middle-east countries, \ie Egypt (EG), Iran (IR), and Israel (IS), for the period from 2015-02-19 to 2022-03-17. Then we download the event news articles with valid URLs, excluding those news articles of which the URLs are broken or inaccessible. To improve the article quality, we further applied document filtering steps by keeping popular and reliable news agencies and entities based on their frequency descending order. This step can largely reduce the amount of low-quality news articles, which are with poor writing or even fake news. Finally, we keep 275,406 documents which is about 47\% of all successfully downloaded news articles.

\subsubsection{Structured Event Extraction} Even though GDELT offers the extracted structured atomic events, unfortunately, such extraction results are prone to coarse-grained events and actors due to its outdated rule-based extraction system built more than ten years ago. Therefore, it is imperative to apply cutting-edge extraction techniques to re-extract the events. Most of the existing EE studies~\cite{ACE2005,DocEE,OneIE} are supervised methods, which require high-quality human-annotated datasets that are labor-intensive and expensive. With the striking success of ChatGPT~\footnote{\url{https://chat.openai.com/}} and GPT-4~\cite{GPT4}, zero-shot information extraction without annotations becomes feasible~\cite{zeroShotIE,LLM4KGconstruction} which demonstrates high potential. We thus leverage LLMs for EE in a zero-shot paradigm. There are tens of optional commercial or open-source LLMs emerging and fast evolving, we take the well-performing open-sourced Vicuna-13b~\footnote{\url{https://chat.lmsys.org/} by May, 2023}, because it is affordable for general academic labs \wrt computational resources and time considering million-level corpus.

\begin{table*}[t]
\caption{The detailed dataset statistics of MidEast-TE.}
\vspace{-0.1in}
\label{tab:clustering_stats}
\centering
\setlength{\tabcolsep}{1mm}{
    \resizebox{0.7\textwidth}{!}{
        \begin{tabular}{l c c c c ccc ccc}
        \toprule
        \multirow{2.4}{*}{$\lambda$} & \multirow{2.4}{*}{\shortstack{minimum \\ cluster size}} & \multirow{2.4}{*}{\#clusters} & \multirow{2.4}{*}{\%outlier doc} & \multirow{2.4}{*}{\shortstack{\%outlier \\ atomic events}} & \multicolumn{3}{c}{\#atomic events} & \multicolumn{3}{c}{\#time span (days)} \\
        \cmidrule(lr){6-8} \cmidrule(lr){9-11} & & & & & max & min & avr & max & min & avr \\
        \midrule
        0 & 10 & 1,653 & 53.93 & 54.60 & 3,257  & 10 & 145.82 & 2,583 & 2  & 1682.16 \\
        1 & 5  & 9,126 & 46.13 & 46.82 & 1,318  & 5  & 30.96  & 1,018 & 1  & 22.74   \\
        \underline{1} & \underline{10} & \underline{3,750} & \underline{52.52} & \underline{52.52} & \underline{4,541}  & \underline{10} & \underline{67.27}  & \underline{2,568} & \underline{2}  & \underline{39.25}   \\
        1 & 30 & 973  & 60.19 & 59.33 & 10,342 & 34 & 222.08 & 2,568 & 2  & 85.53   \\
        1 & 50 & 525  & 63.18 & 62.31 & 10,268 & 55 & 381.38 & 2,567 & 11 & 114.18  \\
        \bottomrule
        \end{tabular}
    }
}
\vspace{-0.1in}
\end{table*}

\begin{table*}[t]
\caption{The detailed statistics of MidEast-TE and GDELT-TE.}
\vspace{-0.1in}
\label{tab:detailed_stats_mideast}
\centering
\setlength{\tabcolsep}{1mm}{
    \resizebox{0.7\textwidth}{!}{
        \begin{tabular}{l l c c c c c ccc ccc}
        \toprule
        \multirow{2.4}{*}{Dataset} & \multirow{2.4}{*}{subset} & \multirow{2.4}{*}{\#docs} & \multirow{2.4}{*}{\shortstack{\#ttl atomic \\ events}} & \multirow{2.2}{*}{$|\mathcal{E}|$} & \multirow{2.4}{*}{$|\mathcal{R}|$} & \multirow{2.4}{*}{\#CEs} & \multicolumn{3}{c}{\#atomic events/CE} & \multicolumn{3}{c}{\#time span (days)} \\
        \cmidrule(lr){8-10} \cmidrule(lr){11-13} & & & & & & & max & min & avr & max & min & avr \\
        \midrule
        \multirow{5.4}{*}{MidEast-TE}
        & train   & 90,252  & 160,953 & 2,641 & 218 & 3,452 & 112 & 10 & 46.63 & 78 & 2 & 30.11 \\
        & val     & 13,532  & 25,401  & 1,281 & 176 & 492   & 112 & 10 & 51.53 & 78 & 2 & 34.33 \\ 
        & test    & 12,333  & 22,440  & 1,183 & 177 & 453   & 112 & 10 & 49.54 & 78 & 2 & 34.92 \\
        & outlier & 137,802 & 247,083 & 2,760 & 227 & -     & -   & -  & -     & -  & - & -     \\
        & total   & 253,836 & 455,877 & 2,794 & 234 & 4,397 & 112 & 10 & 47.49 & 78 & 2 & 31.08 \\
        \midrule
        \multirow{5.4}{*}{GDELT-TE}
        & train   & 101,487 & 441,120   & 1,546 & 233 & 3,561 & 306 & 10 & 123.88 & 78 & 2 & 29.82 \\
        & val     & 14,867  & 66,785    & 1,127 & 207 & 500   & 306 & 10 & 133.57 & 78 & 2 & 34.55 \\ 
        & test    & 13,302  & 65,433    & 1,122 & 204 & 474   & 306 & 10 & 138.04 & 78 & 2 & 34.85 \\
        & outlier & 144,352 & 628,543   & 1,553 & 231 & -     & -   & -  & -      & -  & - & -     \\
        & total   & 273,845 & 1,201,881 & 1,555 & 239 & 4,535 & 306 & 10 & 126.43 & 78 & 2 & 30.87 \\
        \bottomrule
        \end{tabular}
    }
}
\vspace{-0.1in}
\end{table*}

\textbf{Hierarchical Extraction.} One major problem faced by LLM-based EE falls in the input length limitation, especially since the number of distinct atomic event types is over 200 and needs to be input together with necessary extraction instructions and source news articles. To solve this problem, we propose a hierarchical extraction pipeline based on the three-level relation hierarchy in CAEMO from coarse-grain to fine-grain. As shown in the bottom of Figure~\ref{fig:pipeline}, we first input the news article and the first-level atomic event types, prompting the Vicuna to extract all the first-level events, each including event subject, object, and one of the first-level relations. We use the publish date of the news as the timestamp of the extracted events. Then, we parse the first-level extraction results. For each valid first-level event, we input its affiliated second-level relations as options to the Vicuna model, together with the original news article, thus obtaining the second-level extraction results. The same procedure applies to the third-level extraction. Note that we provide a `\textit{No specific}' choice for the sub-level relation extraction, and in this case, the prior level relation will be kept.

\textbf{Entity Linking.} Since we do not have a pre-defined entity set during EE, the extracted entities have free forms and multiple entities correspond to the same one. For example, the extracted entities \textit{U.S.A} and \textit{United States} actually refer to the same actual entity. We address this problem by performing entity linking, which is also a typical step in previous TCE construction studies~\cite{theFuture,resin}. We use the current SOTA LLM GPT-4 to conduct the entity linking since the entity set before linking is relatively small, thus the cost of using GPT-4 is marginal. Specifically, we first apply a K-means clustering over all the original entities to group them into multiple groups. Then we input the entities of each cluster as one batch and ask the GPT-4 to perform entity linking. We have 6322 different entities before entity linking, and we take K to be 32 and obtain a final entity set of size 2794. After the hierarchical extraction and entity linking, we finally obtain all atomic events to construct our dataset MidEast-TE. Even though the EE results in the original GDELT are noisy, it is still worthwhile to use it as an auxiliary dataset. Therefore, we keep the original EE results of GDELT, thus obtaining all atomic events for GDELT-TE. 

\begin{figure*}[t]
    \centering
    \includegraphics[width = 0.98\linewidth]{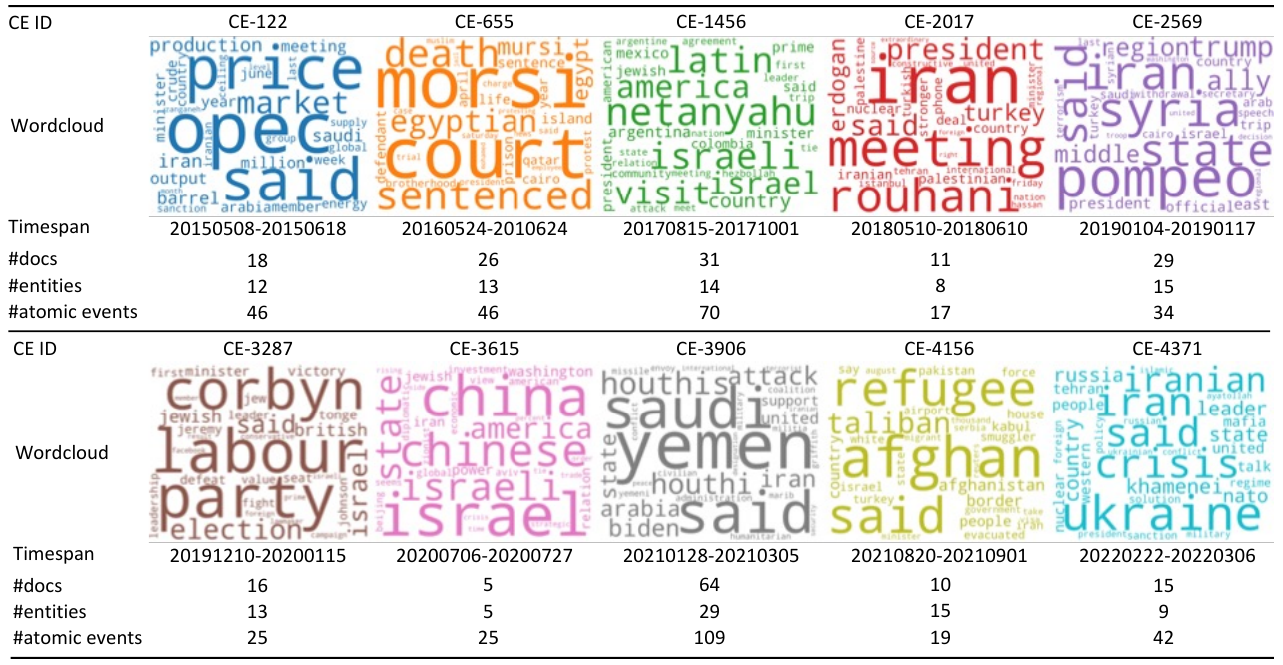}
    \vspace{-0.1in}
    \caption{Illustration of ten examples of CEs, each with its CE ID, wordcloud, timespan, number of involved news articles, entities, and atomic events.}
    \label{fig:example_CEs_wordcloud}
\end{figure*}

\subsubsection{Complex Event Identification} Previous approaches to constructing CEs typically require CE schema construction~\cite{theFuture}, \aka schema induction, which is non-trivial and often requires domain knowledge and extensive iterative work between human and machine~\cite{resin,harvestingEventSchemaLLM}. Recently, several efforts~\cite{scstory,eventStoryLine} about online story discovery employ unsupervised clustering to automatically detect temporally evolving storylines, which is simpler and more flexible without the need to construct any CE schemas. 

Following these approaches, we develop a time-aware document clustering pipeline as shown in the top of Figure~\ref{fig:pipeline}. Specifically, we first use a SimCSE~\cite{SimCSE} pre-trained RoBERTa~\cite{RoBERTa} to extract the document embedding. Then, we follow BERTopic~\cite{BertTopic} that uses UMAP~\cite{UMAP} to reduce the dimensionality of embeddings and HDBSCAN~\cite{HDBSCAN} to cluster the documents into semantically related groups. In addition to semantic similarity, news articles within the same CE should also be temporally close to each other. Previous works~\cite{scstory} use a sliding window with a fixed length to split the timeline into pieces, which may not be optimal since different TEs often have varied time spans. To resolve this limitation, we propose a time-aware clustering approach that concatenates the temporal index of the news article to its semantic embedding, thus endowing the article clusters with elastic time spans. Moreover, we introduce an additional hyper-parameter $\lambda$ to balance the weight of temporal affinity against the semantic similarity during clustering. After the clustering, we obtain a set of document clusters $\mathbf{D}^c$, each representing a CE. Apart from the CEs, there are quite several isolated news articles that do not belong to any cluster. We deem that they still offer precious global contextual information that is helpful to TE forecasting, so we keep the atomic events extracted from these documents as \textit{outlier atomic events} in the global context.

Table~\ref{tab:clustering_stats} shows our parameter tuning in the clustering process. Specifically, after we reduce document semantic embedding to 200 using UMAP, we tune the hyper-parameter of \textit{minimum cluster size} in HDBSCAN from $\{5,10,30,50\}$, and tune the temporal feature weight $\lambda$ from $\{0,1\}$. We observe that when $\lambda=0$, which means when the clustering does not consider the temporal feature and solely relies on the semantic features, the average time span of the cluster (CE) is extremely large as 1682 days (4.6 years). In contrast, under the setting of $\lambda=1$, the average CE time span is largely reduced, showing the effectiveness of the time feature in constraining the temporal conciseness of CEs. We empirically take 10, which will result in a reasonable scale for the average cluster size, \wrt both \#atomic events (67.27) and \#time span (39.25). 

After Clustering, there are a few extremely large clusters, for example, the max cluster has 4541 atomic events and spans 2568 days. We divide such superclusters into multiple reasonable smaller clusters by empirically setting a maximum \#atomic events $h_a$ and a maximum \#time span $h_t$ to twice the original average. Then we remove extremely small clusters that are shorter than 2 days or have fewer than 10 atomic events. We then split CEs into training/validation/testing sets in time order to prevent information leakage in forecasting. We apply this entire pipeline to the atomic events obtained from our LLM-based EE and from GDELT. We finally obtain two large-scale SCTc-TE datasets: MidEast-TE and GDELT-TE, with the detailed statistics as shown in Table~\ref{tab:detailed_stats_mideast}.

\subsection{Dataset Evaluation} 

\subsubsection{Statistics} 
\begin{table}[t]
\begin{center}
\caption{Dataset comparison between our SCTc-TE and TCEs with schema.}
\label{tab:dataset_1}
\vspace{-0.1in}
\resizebox{0.45\textwidth}{!}{
    \begin{tabular}{ll ccc}
        \toprule
        Formulation & Dataset & \#docs & \#CEs & \#atomic events \\
        \midrule
        \multirow{2}{*}{TCE w Schema}  &  General     & 617     & 617     & 8,295 \\
                                       & IED          & 6,399   & 430     & 51,422 \\
        \midrule
        \multirow{2}{*}{\textbf{SCTc-TE} (ours)}  & GDELT-TE    & 273,845 &  4,397  & 1,201,881 \\
                                  & MidEast-TE  & 274,795 &  4,397  & 455,877 \\
        \bottomrule
    \end{tabular}
}
\end{center}
\vspace{-0.1in}
\end{table}

Table~\ref{tab:dataset_1} shows the dataset comparison between the previous formulation of TCE with Schema and our formulation of SCTc-TE. The \textbf{General} dataset is curated by LDC (LDC2020E25) and the \textbf{IED} is constructed by~\citet{theFuture}. Clearly, the two datasets constructed by our pipeline are much larger than the previous datasets; in particular, the number of CEs and atomic events are one magnitude larger than the previous datasets.

\begin{table}[t]
\begin{center}
\vspace{-0.1in}
\caption{Event extraction results comparison.}
\label{tab:dataset_2}
\vspace{-0.1in}
\resizebox{0.45\textwidth}{!}{
    \begin{tabular}{lcccccc}
        \toprule
        \multirow{2.4}{*}{Dataset} & \multirow{2.4}{*}{| $\mathcal{R}$ |} & \multirow{2.4}{*}{| $\mathcal{E}$ |} & \multirow{2.4}{*}{\shortstack{\#atomic \\ events}} & \multicolumn{3}{c}{\% of different levels} \\
        \cmidrule(lr){5-7} & & & & 1st & 2nd & 3rd \\
        \midrule
        GDELT-TE   & 239 & 1,555 & 1,201,881 & 38.93 & 57.19 & 3.88  \\
        MidEast-TE & 234 & 2,794 & 455,877   & 21.24 & 64.82 & 13.94  \\
        \bottomrule
    \end{tabular}
}
\end{center}
\end{table}

\begin{table}[t]
\begin{center}
\caption{Dataset Statistics.}
\label{tab:dataset_3}
\resizebox{0.45\textwidth}{!}{
    \begin{tabular}{ll ccc}
        \toprule
        Dataset & specs & train & val & test \\
        \midrule
        \multirow{2}{*}{GDELT-TE}   & \#CEs          & 3,561   & 500    & 474 \\
                                    & \#atomic events & 441,120 & 66,785 & 65,433 \\
        \midrule
        \multirow{2}{*}{MidEast-TE} & \#CEs          & 3,452   & 492    & 453 \\
                                    & \#atomic events & 160,953 & 25,401 & 22,440 \\
        \bottomrule
    \end{tabular}
}
\end{center}
\end{table}

We compare MidEast-TE and GDELT-TE in Table~\ref{tab:dataset_2}. The events in MidEast-TE are more fine-grained due to a larger entity set. Moreover, the atomic event types defined by CAMEO are a three-layered hierarchy, and the event types at higher levels are more fine-grained. The distribution of events in different levels verifies that the events in MidEast-TE are more fine-grained. We sample 100 news articles and use GPT-3.5 to evaluate the EE performance, where the precison of MidEast-TE and GDELT-TE are 0.432 and 0.425, respectively. Event though the EE using Vicuna-13b is a little bit better than the GDELT, the overall error rate in current EE of both MidEast-TE and GDELT-TE are still high. However, we argue that the idea of using LLMs for EE is promising. If there is no budget limitation and we use more powerful LLMs, the error rate of EE could be largely reduced. We split the CEs into train/validation/testing sets according to the timestamp. Specifically, we use the CEs in the last year for testing, the second-to-last year for validation, and the remaining of about five years for training, as shown in Table~\ref{tab:dataset_3}.

\subsubsection{Evaluation}
We specifically evaluate the three main steps of the dataset construction process, including clustering, event extraction, and entity linking. For the clustering, we sample some of them and plot their word clouds using tf-idf scores. The examples in Figure~\ref{fig:example_CEs_wordcloud} show that the clusters are distinct with each other, and each of the cluster focus on specific complex event. For the event extraction, we adopt more powerful commercial LLMs, \ie gpt-3.5 (which is more efficient and cheap compared with the most powerful GPT-4), to evaluate the extraction precision of both MidEast-TE and the GDELT-TE (the original EE system of GDELT). We sample 100 news articles from the entire datasets and input them as well as the extracted events into the prompt, then we ask the LLMs whether the extraction results are correct or wrong. The results show that the precision of MidEast-TE is 0.432 and the precision of GDELT-TE is 0.425. Overall speaking, the extraction performances of both LLMs and GDELT are still quite low. Nevertheless, our LLM-based EE is slightly better than the original GDELT. The open sourced Vicuna-13b verifies that it is possible to be used as a zero-shot EE model. In the future, we believe the performance of EE will be largely improved when more efficient and powerful LLMs are developed. For the entity linking, since the remaining entity set is small (2794), we manually check and fix the entity linking results, ensuring the linking result is correct.
\begin{figure*}[t]
    \centering
    \includegraphics[width = 0.88\linewidth]{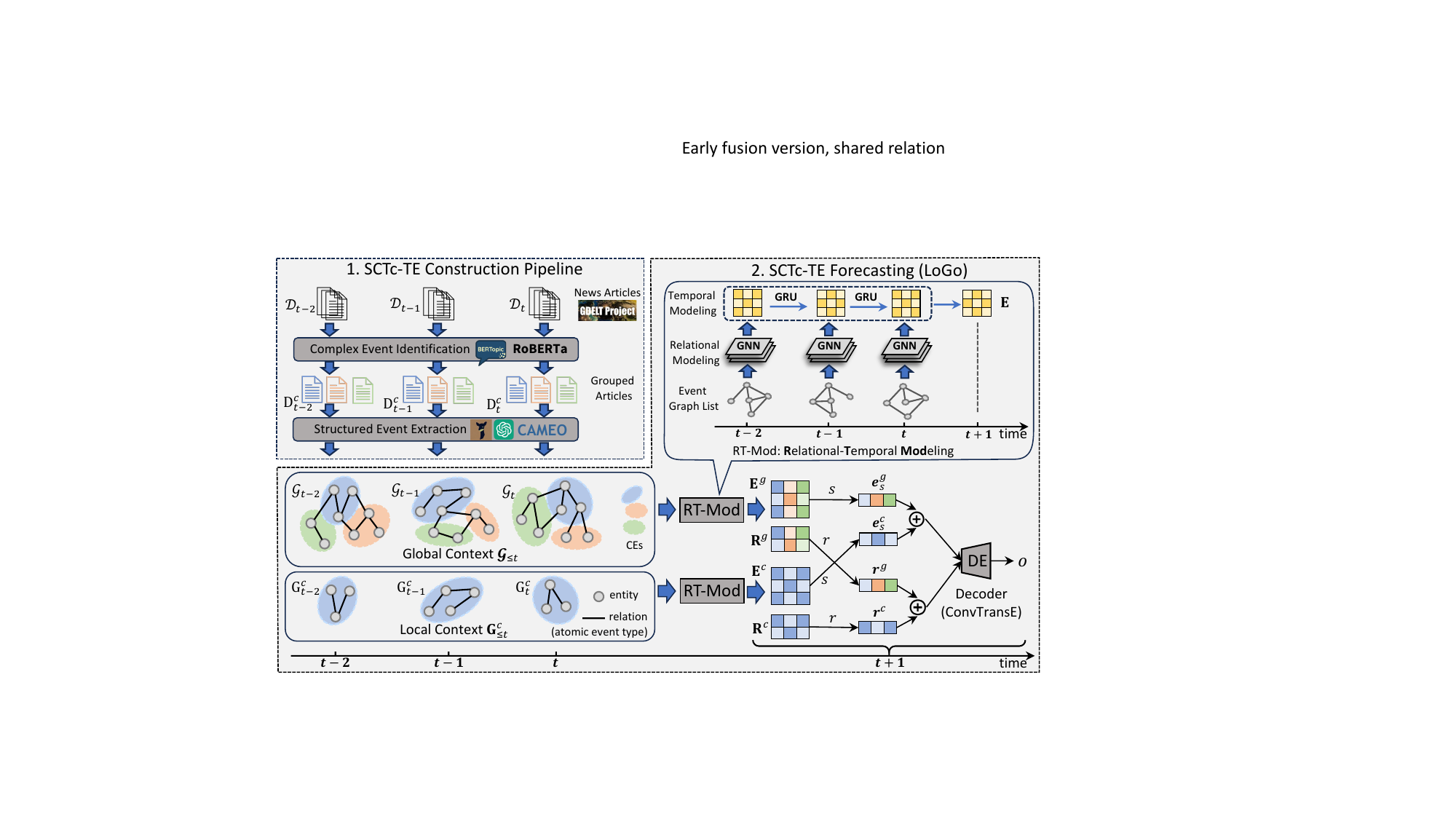}
    \vspace{-0.1in}
    \caption{The overall framework of this work consists of two parts: 1) the SCTc-TE construction pipeline (the top left part), and 2) the SCTc-TE forecasting model LoGo (the right and bottom part).}
    \vspace{-0.1in}
    \label{fig:framework}
\end{figure*}

\section{Methodology} \label{sec:method}

Third, we propose a novel TE forecasting method, \aka LoGo, as shown in the bottom and right of Figure~\ref{fig:framework}. 

\subsection{SCT Event Forecasting} \label{subsec:tce_forecasting}

Our new SCTc-TE formulation is well suited for the downstream forecasting task. The structured and time-complete properties enable SCTc-TE forecasting to be modeled by most existing TKG methods. However, previous TKG methods mix up all the atomic events and do not distinguish among different CEs, resulting in sub-optimal modeling capability. In this work, we design a novel approach named LoGo that leverages both \textbf{Lo}cal and \textbf{G}l\textbf{o}bal contexts for event forecasting. Specifically, given a specific query $(s, r, ?, t+1, c)$, we first obtain the entity and relation representations through the local and global context modeling modules, we then fuse these representations learned from both contexts, and finally, we use the decoder ConvTransE to predict the object $o$. The overall forecasting framework is shown in Figure~\ref{fig:framework}.  

\subsubsection{Local and Global Context Modeling} Following the previous work RE-GCN~\cite{REGCN}, we design a simple module Relational-Temporal Modeling, named as RT-Mod, for the local and global context modeling. We employ a GNN model, \ie RGCN~\cite{RGCN}, to capture the relations among concurrent events at the same timestamp and an RNN model, \ie GRU~\cite{GRU}, to preserve the temporal evolving patterns along the timeline. 

\textbf{Local Context Modeling.} Based on the $G^c_t$ of CE $c$ at timestamp $t$, the information propagated to entity $o$ via RGCN is $\mathbf{e}^{l}_{o,c,t} \in \mathbb{R}^d$, represented as: 
\begin{equation} \label{eq_1}
    \mathbf{e}^{l}_{o} = f \Bigl( \frac{1}{\vert G^c_t \rvert} \sum_{(s,r,o) \in G^c_t}{\mathbf{W}^l_1(\mathbf{e}^{l-1}_{s} + \mathbf{r}) + \mathbf{W}^{l}_{2} \mathbf{e}^{l-1}_{o}} \Bigr), 
\end{equation}
where $d$ is the dimensionality of the propagated information, $\vert G^c_t \rvert$ represents the number of events where $o$ is the object, $\mathbf{W}^l_1, \mathbf{W}^l_2 \in \mathbb{R}^{d \times d}$ are the parameters of the GNN kernel for layer $l$, and $f(\cdot)$ is the RReLU activation function. It is noted that $\mathbf{e}^{l}_{o}, \mathbf{e}^{l-1}_{s}, \mathbf{r}, \mathbf{e}^{l-1}_{o}$ correspond to their representations in CE $c$ at timestamp $t$, where we omit the subscript $(c,t)$ for simplicity.

Thereafter, we aggregate the information received from multi-layer propagation and obtain the entity representation at time $t$ under context $c$, defined as:
\begin{equation} \label{eq_2}
    \mathbf{e}_{o} = \sum^{L^c}_{l=0}{\mathbf{e}^l_{o}},
\end{equation}
where $\mathbf{e}^0_{o} \in \mathbb{R}^{d}$ is randomly initialized for each entity $o$. Note that the same entity share one initialized embedding over different CEs, while its representation will change differently after the graph propagation on its involved $G_t^c$. $L^c$ is the number of propagation layers, which is set as the same value for all the CEs $\mathcal{C}$. The representations for all entities at timestamp $t$ are denoted as $\mathbf{E}^{c}_{t} \in \mathbb{R}^{| \mathcal{E} | \times d}$.
Then, we apply a GRU to capture the temporal patterns, defined as:
\begin{equation} \label{eq_3}
    \mathbf{E}^{c}_{t} = \text{GRU}^{c}( \mathbf{E}^{c}_{t}, \mathbf{E}^{c}_{t-1}),
\end{equation}
where $\text{GRU}^{c}$ represents the GRU kernel for the local context. Following the typical implementation in TKG studies~\cite{TKG-survey}, we take the recent $T^c$ steps of historical graphs to model the temporal evolving patterns for the local context. Thereby, the entity embeddings in the last GRU state retain the local contextual information for CE $c$, and we denote them as $\mathbf{E}^c$. All the relation embeddings in the local context are denoted as $\mathbf{R}^c \in \mathbb{R}^{| \mathcal{R} | \times d}$.

\textbf{Global Context Modeling.} Analogous to local context modeling, we employ a parallel branch of RT-Mod module to capture the historical patterns in the global context. Given the global context $\mathcal{G}_t$ at timestamp $t$, we obtain the entity representations $\mathbf{E}^{g}_{t} \in \mathbb{R}^{| \mathcal{E} | \times d}$ after $L^g$ layers of RGCN information propagation and aggregation. Thereafter, by going through the GRU module $\text{GRU}^{g}$ over the latest $T^g$ steps in the global context $\mathcal{G}$, we achieve the entity representations $\mathbf{E}^{g} \in \mathbb{R}^{| \mathcal{E} | \times d}$. Note that all the entity and relation embeddings, as well as the RGCN and GRU parameters of the global context are not shared with the local context. In addition, the relation embeddings in the global context are denoted as $\mathbf{R}^g \in \mathbb{R}^{| \mathcal{R} | \times d}$.

\begin{table*}[t]
\caption{The overall performance of our LoGo and baselines. "\%Improv." denotes the relative improvement.}
\label{tab:overall_performance}
\centering
\setlength{\tabcolsep}{1mm}{
    \resizebox{0.9\textwidth}{!}{
        \begin{tabular}{ll cccc cccc cc}
        \toprule
         \multirow{2.4}{*}{Dataset} & \multirow{2.4}{*}{Metric} & \multicolumn{4}{c}{Static KG Methods} & \multicolumn{4}{c}{TKG Methods} & \multicolumn{2}{c}{Ours} \\
         \cmidrule(lr){3-6} \cmidrule(lr){7-10} \cmidrule(lr){11-12} & & DistMult & ConvE & ConvTransE & RGCN & RE-NET & RE-GCN & HisMatch & CMF & \textbf{LoGo} & \%Improv. \\
        \midrule
        \multirow{4}{*}{GDELT-TE}
              & MRR    & 0.1148 & 0.1140 & 0.1204 & 0.1277 & 0.1179 & 0.1403 & \underline{0.1641} & 0.1387 & \textbf{0.2533} & 54.33 \\
              & HIT@1  & 0.0368 & 0.0387 & 0.0422 & 0.0503 & 0.0338 & 0.0507 & \underline{0.0716} & 0.0499 & \textbf{0.1344} & 87.62 \\
              & HIT@3  & 0.1109 & 0.1107 & 0.1191 & 0.1285 & 0.1191 & 0.1379 & \underline{0.1710} & 0.1397 & \textbf{0.2899} & 69.51 \\
              & HIT@10 & 0.2843 & 0.2734 & 0.2847 & 0.2874 & 0.3006 & 0.3445 & \underline{0.3652} & 0.3345 & \textbf{0.5053} & 38.37 \\
        \midrule
        \multirow{4}{*}{MidEast-TE}
              & MRR    & 0.2227 & 0.2276 & 0.2358 & 0.2744 & 0.2504 & 0.2989 & \underline{0.3354} & 0.2857 & \textbf{0.4978} & 48.44 \\
              & HIT@1  & 0.1177 & 0.1172 & 0.1283 & 0.1795 & 0.1447 & 0.1874 & \underline{0.2226} & 0.1722 & \textbf{0.3838} & 72.42 \\
              & HIT@3  & 0.2568 & 0.2674 & 0.2731 & 0.3083 & 0.2833 & 0.3373 & \underline{0.3826} & 0.3249 & \textbf{0.5708} & 49.18 \\
              & HIT@10 & 0.4421 & 0.4634 & 0.4658 & 0.4769 & 0.4740 & 0.5378 & \underline{0.5642} & 0.5214 & \textbf{0.6956} & 23.31 \\
        \bottomrule
        \end{tabular}
    }
}
\end{table*}

\subsubsection{Prediction and Optimization} After the local and global context modeling, we obtain two sets of representations for entities and relations, \ie $[\mathbf{E}^c, \mathbf{R}^c]$ from the local context and $[\mathbf{E}^g, \mathbf{R}^g]$ from the global context. Given a query $(s,r,?,t+1,c)$, we lookup these embedding tables and yield the corresponding representations $[\mathbf{e}^c_s, \mathbf{r}^c]$ and $[\mathbf{e}^g_s, \mathbf{r}^g]$. We perform the element-wise sum operation to combine the representations from both contexts, resulting in the final representation $[\mathbf{\hat{e}}_s, \mathbf{\hat{r}}]$. Then, we resort to ConvTransE~\cite{ConvTransE} as the decoder, formally written as:
\begin{multline} \label{eq_5}
    \mathbf{\hat{p}}(\mathcal{E}|s,r,c,\mathbf{G}^c_{\le t},\mathcal{G}_{\le t})= \text{softmax} \bigl( \mathbf{\hat{E}} \text{ConvTransE}(\mathbf{\hat{e}}_{s}, \mathbf{\hat{r}}) \bigr),
\end{multline}
where $\text{sotfmax}(\cdot)$ is the softmax function, $\text{ConvTransE}(\cdot)$ is the decoder, and $\mathbf{\hat{E}}$ is the candidate entity embedding, which is summed over $\mathbf{E}^c$ and $\mathbf{E}^g$. The prediction is denoted as: 
\begin{equation} \label{eq_6}
    \hat{o}_{(s,r,t+1,c)} = \arg\max_{\mathcal{E}}{\hat{p}(\mathcal{E}|s,r,c,\mathbf{G}^c_{\le t},\mathcal{G}_{\le t})}.
\end{equation}
It should be noted that our implementation is a type of early fusion that we first fuse the representations from two branches and then feed them into the decoder. This early fusion strategy can fully take advantage of the power of the decoder ConvTransE, which is a multi-layer convolutional neural network, to capture the interactions between the two contexts. In this way, the model can adaptively learn an optimal combination of the two contextual representations to yield the accurate prediction for a certain query. We utilize the cross-entropy loss defined below to optimize our model:
\begin{multline} \label{eq_7}
    \mathcal{L} = \sum_{c \in \mathcal{C}} \sum_{(s,r) \in G^c_{t+1}} \mathbf{y}_{(s,r,t+1,c)} \log \mathbf{\hat{p}}(\mathcal{E}|s,r,c,\mathbf{G}^c_{\le t},\mathcal{G}_{\le t}).
\end{multline}
\section{Experiments} \label{sec:experiments}

We conduct experiments on the two SCTc-TE datasets MidEast-TE and GDELT-TE to justify the proposed model. We are particularly interested in answering these research questions:

\begin{itemize}[leftmargin=*]
    \item \textbf{RQ1: } Does our proposed LoGo outperform the SOTA methods, especially the TKG methods?
    \item \textbf{RQ2: } How do the key model designs contribute to the forecasting?
    \item \textbf{RQ3: } What are the effects of various hyper-parameter settings and how the model perform in concrete cases?
\end{itemize}

\subsection{Experimental Settings}
Following the typical settings of TKG~\cite{TKG-survey}, we employ the evaluation metrics of Mean Reciprocal Rank (MRR) and Hit Rate(HIT)@\{1, 3, 10\} and use time-aware filtering during testing. The MRR is used to select the best-performing models based on the validation set. The implementation details are described in Appendix~\ref{sec:appendix}.

\subsubsection{Compared Methods}
We compare our method with two strands of baseline methods, \ie static KG methods and TKG methods, which are widely used in the event forecasting problem under the TKG formulation. It should be noted that the methods designed for TCE with schema do not apply to our SCTc-TE problem since we do not have a pre-defined CE schema. 
\textbf{Static KG Methods} ignore the timestamp of the atomic event and combine all the triplet-formatted atomic events. We consider several representative methods. DistMulti~\cite{DistMult}, ConvE~\cite{ConvE}, ConvTransE~\cite{ConvTransE}, and RGCN~\cite{RGCN}.
\textbf{TKG Methods} are designed for temporal atomic events forecasting. Specifically, we encompass several representative models, including RE-NET~\cite{RENET}, RE-GCN~\cite{REGCN}, HisMatch~\cite{HiSMatch}, and CMF~\cite{CMF}. HisMatch is the SOTA method. CMF can utilize the content information (news article embedding) of historical atomic events.

\subsubsection{Implementation Details}
We implement all the static methods by ourselves. In terms of RE-NET, RE-GCN, and HisMatch, we reuse their officially released code and use the global context graph. We re-implement CMF by ourselves according to the original paper, and the textual embedding is extracted using the RoBERTa model fine-tuned by SimCSE. For all the baseline and our methods, only the atomic events within CEs are used as training samples and predicted during testing. For the outlier atomic events, they are just used as part of the input graph. To be fair, for all the methods, set the embedding dimensionality $d=200$, apply cross-entropy loss, grid search learning rate from $\{10^{-2}, 10^{-3}, 10^{-4}\}$ and weight decay from $\{10^{-4}, 10^{-5}, 10^{-6}, 10^{-7}\}$. We search the historical length $D$ from $\{1,3,5,7,10,14\}$ for all the TKG methods. The number of propagation layers of all the graph-based methods is grid searched from $\{1,2,3\}$. We use Adam~\cite{Adam} optimizer and Xavier~\cite{Xavier} initialization for all the methods.


\begin{figure*}
    \centering
    \includegraphics[width = 0.95\linewidth]{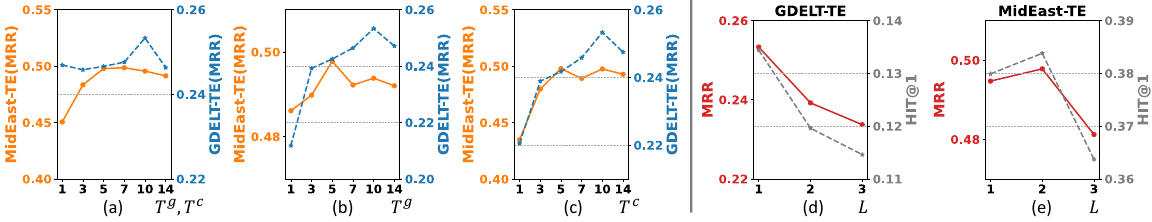}
    \caption{Sub-figure (a,b,c) show the effects of context historical length, and (d,e) study the effects of GNN layers.}
    \label{fig:model_study}
\end{figure*}

\subsection{Performance Comparison (RQ1)}
Table~\ref{tab:overall_performance} presents the overall performance of our method and all the baselines. We have the following observations from the results. First, our method outperforms all the baselines of both static KG and SOTA TKG methods by a large margin, demonstrating the effectiveness of our method. Second, among the two types of baselines, TKG methods generally perform better than the static KG methods, indicating that there are salient temporal patterns within our datasets and properly capturing the temporal patterns is crucial for effective forecasting. Third, graph-based methods generally perform well, showing that GNN is still the leading technique for structured temporal event forecasting. Moreover, the text-based method CMF does not outperform purely structured methods, showing that there is still a large space to explore for incorporating textual information into event forecasting.

\begin{table}[t]
\begin{center}
\caption{Ablation Study.}
\label{tab:ablation_study}
\resizebox{0.46\textwidth}{!}{
    \begin{tabular}{llcccc}
        \toprule
        Dataset & Model & MRR & HIT@1 & HIT@3 & HIT@10 \\
        \midrule
        \multirow{5}{*}{GDELT-TE} & $\text{Logo}_\text{local}$  & 0.2164 & 0.1067 & 0.2490 & 0.4460 \\ 
                                    & $\text{Logo}_\text{global}$ & 0.1471 & 0.0539 & 0.1517 & 0.3509 \\
               & $\text{Logo}_\text{share}$  & 0.2391 & 0.1200 & 0.2734 & 0.4940 \\
               & $\text{Logo}_\text{late}$   & 0.2082 & 0.1050 & 0.2328 & 0.4271 \\
               & LoGo    & 0.2414 & 0.1224 & 0.2759 & 0.4970 \\
        \midrule
        \multirow{5}{*}{MidEast-TE} & $\text{Logo}_\text{local}$  & 0.4230 & 0.3258 & 0.4784 & 0.6088 \\ 
                                    & $\text{Logo}_\text{global}$ & 0.2920 & 0.1765 & 0.3317 & 0.5367 \\
               & $\text{Logo}_\text{share}$  & 0.4850 & 0.3726 & 0.5570 & 0.6810 \\
               & $\text{Logo}_\text{late}$  & 0.4018 & 0.3080 & 0.4590 & 0.5720 \\
               & LoGo   & 0.4978 & 0.3838 & 0.5708 & 0.6956 \\
        \bottomrule
    \end{tabular}
}
\end{center}
\end{table}

\subsection{Ablation Study (RQ2)}
we design several ablated models to justify the key designs of LoGo. First, we just keep one of the two contexts and set up two model variants, \ie $\text{Logo}_\text{local}$ that only keeps the local context branch and $\text{Logo}_\text{global}$ that keeps the global context branch. The results on two datasets in Table~\ref{tab:ablation_study} indicate that: 1) solely relying on either of the local and global contexts will result in a performance drop; and 2) between the two individual contexts, the local context is more valuable than the global context since $\text{Logo}_\text{local}$ outperforms $\text{Logo}_\text{global}$ on both datasets of all the metrics. This makes sense because the local context zoom in to the evolution of the specific CE, thus making the model focus more on the entities that appear in the CE. Nevertheless, the global context can provide auxiliary environmental information for those sparse entities that have few information in the local context, which is the reason why LoGo is more effective than $\text{Logo}_\text{local}$. 

Second, we design $\text{Logo}_\text{share}$ that let the two contexts share parameters.
The performance of $\text{Logo}_\text{share}$ is a little bit lower than that of LoGo, demonstrating that two separate branches with isolated parameters can better capture the characteristics in both contexts. However, the performance of $\text{Logo}_\text{share}$ is still much better than other baselines.  

Third, we try late fusion strategy of combining the two contexts, \ie $\text{Logo}_\text{late}$. The default strategy of LoGo is early fusion, while $\text{Logo}_\text{late}$ lets the two branches first pass two separate decoders and then fuse the decoded representations for ranking. The results in Table~\ref{tab:ablation_study} show that early fusion is superior to late fusion. It verifies our motivation that the convolutional layers in the following decoder (ConvTransE) can well capture the interactions of the early-fused representations for optimal forecasting. 


\subsection{Model Study (RQ3)}
\textbf{Effect of Historical Length.} We vary the historical length of $T^g$ and $T^c$, and the MRR curve is shown in Figure~\ref{fig:model_study} (a), which illustrates that 5 is the best for MidEast-TE and 10 is the best for GDELT-TE. Then we keep one of $T^g$ and $T^c$ as the best setting and change the other one and show the MRR curves in Figure~\ref{fig:model_study} (b,c). The results indicate that proper historical length is crucial for good performance.


\textbf{Effect of Propagation Layers.} We vary the number of propagation layers of RGCN and demonstrate the performance change in Figure~\ref{fig:model_study} (d,e). MidEast-TE performs best when $L=2$, while $L=1$ is best for GDELT-TE, showing that MidEast-TE is more sensitive to higher-order connections.

\begin{figure}
    \centering
    \includegraphics[width = 0.99\linewidth]{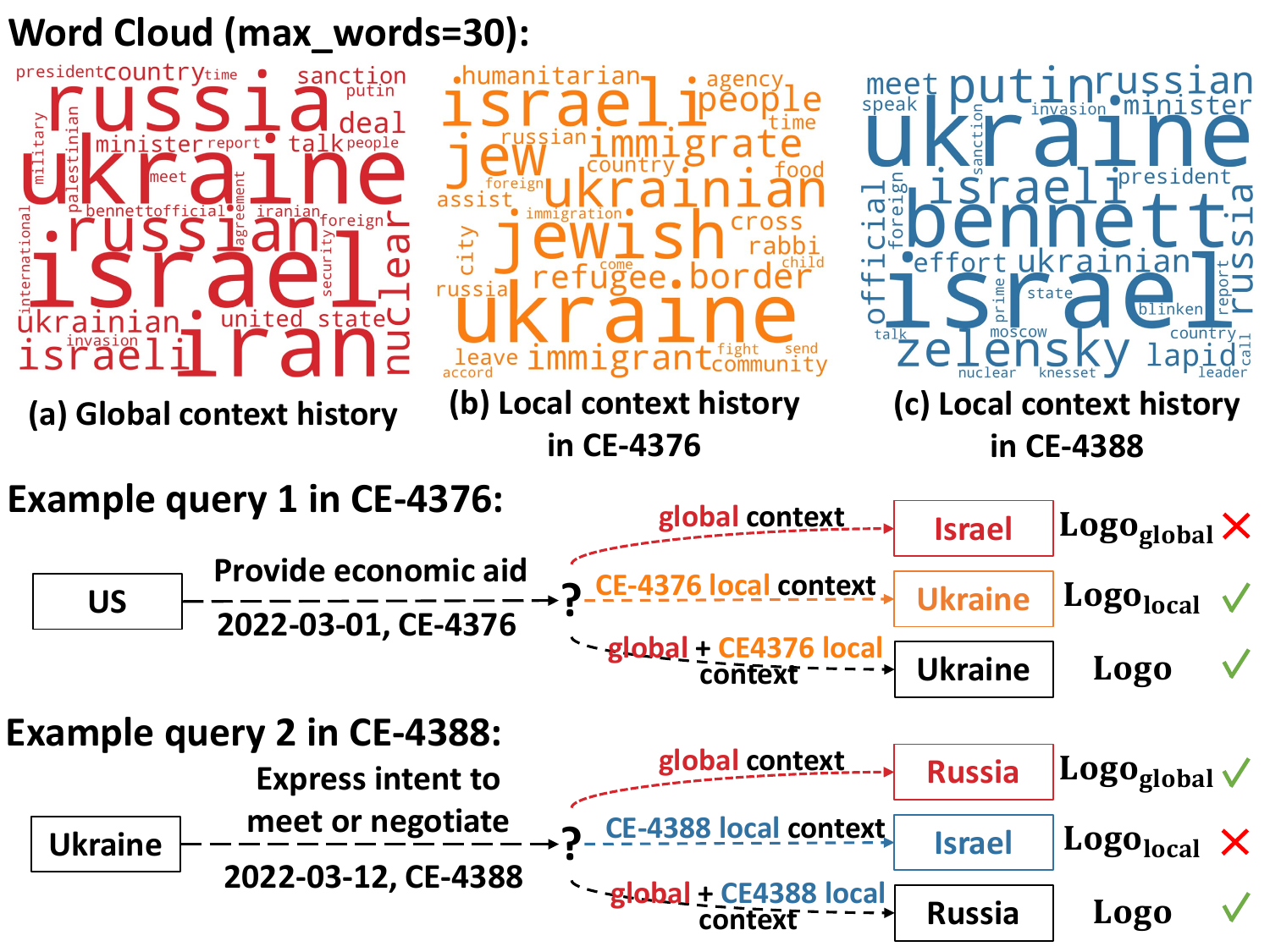}
    \caption{Example queries, corresponding news word clouds, and forecasting results from MidEast-TE.}
    \label{fig:case_study}
\end{figure}

\textbf{Case Study.}
As shown in Figure \ref{fig:case_study}, for each query $(s, r, ?, t+1, c)$, we list the object $o$ predicted by $\text{Logo}$, $\text{Logo}_\text{global}$, and $\text{Logo}_\text{local}$. We plot the word clouds based on tf-idf scores of the news articles in local and global context, \ie $\mathbf{D}^c_{\le t}$ and $\mathcal{D}_{\le t}$.
In query 1~\footnote{Here is the \href{https://www.timesofisrael.com/liveblog_entry/us-orthodox-leader-says-community-offering-homes-funding-for-ukraine-refugees/}{source url} of the referred news article.}, $\text{Logo}_\text{global}$ wrongly predicts that \textit{US} would provide economic aid to \textit{Israel}, but $\text{Logo}_\text{local}$ and $\text{Logo}$ correctly predict the true object \textit{Ukraine} by incorporating the local context of \textit{CE-4376}. From word clouds (a) and (b), we could observe that the global context involves a much larger number of international events and major countries, while \textit{CE-4376} mainly depicts events about Ukraine refugees and the aid from other parties. There are indeed many positive past interactions between \textit{US} and \textit{Israel}, but the local context makes the model focus on the Ukraine refugee problem which leads to a better prediction. This demonstrates that the local context information is crucial. 
On another hand, in query 2~\footnote{Here is the  \href{https://www.timesofisrael.com/zelensky-suggests-jerusalem-host-negotiations-between-ukraine-russia/}{source url} of the referred news article.},  $\text{Logo}_\text{global}$ and $\text{Logo}$ correctly predict that \textit{Ukraine} would express intent to negotiate with \textit{Russia} while $\text{Logo}_\text{local}$ wrongly predict to \textit{Israel}. One possible reason is that \textit{CE-4388} mainly focuses on how Israel was interacting with Ukraine as shown in word cloud (c), but the global context contains more related past events, and thus provides a larger picture of the Russia-Ukraine conflict and its progress. This shows the importance of global context information.
\section{Related Work} \label{sec:related_work}


\textbf{Event Extraction.} Typical EE tasks include sentence-level EE~\cite{ACE2005} and more complex document-level EE~\cite{DocEE}. However, all of these works focus on small-scale individual events while ignoring dependencies between events including their temporal ordering.
Recently, several studies have been conducted for complex event-level EE. One line of work focuses on structured CE extraction, such as IED~\cite{theFuture} and RESIN-11~\cite{resin}. They first design some complex event schema, and then extract complex events that follow such schema. However, such extraction is highly dependent on schema construction, which is a non-trivial step and difficult to scale up to various CE types. Another line of work explores unstructured complex event analysis in a formulation of storyline discovery from multiple documents~\cite{scstory,eventStoryLine}. They employ unsupervised clustering methods to construct storylines from a large set of news corpus, getting rid of schema construction. Therefore, in this work, we follow this clustering-based approach to identify complex events without a rigid schema, and we further extract out the structured events from the unstructured news text for downstream analysis and forecasting.
A more recent work~\cite{smartbook} generates natural language report analysis for temporal complex events using LLMs, but lacks quantitative analysis of the generated report. With the recent explosive progress of LLMs, researchers also take advantage of their zero-shot or few-shot capabilities to serve as a structured event extractor~\cite{eventSchemaInduction,harvestingEventSchemaLLM}. Therefore, We also leverage LLMs~\cite{vicuna2023,GPT4} for the EE in a zero-shot paradigm.


\textbf{Temporal Event Forecasting.}
Temporal event forecasting aims to predict future events at future timestamps based on observed historical events. Many efforts have been devoted to TE forecasting~\cite{survey-event-forecasting} and different formulations have been raised. One line of work represents TE with time series~\cite{timeSeriesEvent,benjamin2023hybrid,RCT_B}, but they fail to capture the multi-relational relation between entities, not to say to model complex events. Several studies also explore the unstructured textual formulation for TE, where a CE is a storyline~\cite{storyLine}, and a TE is either in the form of summaries~\cite{gholipour-ghalandari-etal-2020-large} or phrases~\cite{eventChain}. However, redundant information is largely introduced due to the natural language expression and hinders the downstream forecasting task formulation. The benchmark ForecastQA~\cite{forecastqa} also falls in this unstructured formulation by providing only the textual historical events during forecasting, but the construction of this QA dataset requires heavy human labeling and it only includes TE-level questions. In contrast, structured TE formulations mainly include TKG~\cite{TKG-survey} and TCE with schema~\cite{theFuture,schemaGuidedEvent}. However, TKG limits event analysis and forecasting at the TE level and therefore fails to capture multiple actors, relations, and timelines at the CE level. While TCE with schema models both CE and TE, its dependency on schema induction leads to unflexible event representation and restricts TEs to a relative temporal ordering instead of having complete timestamps. To the best of our knowledge, none of the previous works satisfies all the three properties of structured, complex, and time-complete, and in this work, our SCTc-TE formulation highlights these three critical properties for TE and CE forecasting.

Based on our SCTc-TE, the most promising methods are methods proposed for TKG forecasting. Static KG methods~\cite{ConvE, ConvTransE} treat event forecasting as a link prediction task on a static event graph while ignoring event timestamps. TKG-based methods consider temporal information aside from relational information. For example, RE-NET~\cite{RENET}, RE-GCN~\cite{REGCN} and EvoKG~\cite{EvoKG} use GNN to aggregate the relational information at each timestamp and then use RNN to propagate this information over time. Some work also tries to map the discrete timestamps in a continuous space. For example, Know-Evolve~\cite{knowevolve} models event time as the Temporal Point Process (TPP), and TANGO~\cite{TANGO} uses Neural Ordinary Differential Equation (ODE). Efforts have also been made to identify related historical events or a reasoning path to enhance the event forecasting performance. For example, CyGNet~\cite{cygnet} retrieves relevant TE, and CluSTeR~\cite{cluster} and TimeTraveler~\cite{TimeTraveler} search for event evidence chain. Some TKG-based methods also try to incorporate textual information in addition to structural information into the event forecasting model. For example, Glean~\cite{Glean} uses word graph embedding and CMF~\cite{CMF} uses text embedding of the event relation ontology to enrich the information stored in each relational link in TKG. In parallel, multiple efforts have been devoted to schema-guided TCE forecasting~\cite{theFuture,schemaGuidedEvent}, however, such methods operate on a more complicated event graph definition, which cannot be easily adapted to our setting.

\section{Conclusion and Future Work} \label{conclusion}
In this work, we highlighted three key properties of temporal events, structured, complex, and time-complete, and presented a new formulation of SCTc-TE. Thereby, we developed a fully automated pipeline that employs both LLM and time-aware clustering to construct SCTc-TEs from a large amount of news articles, and we constructed two large-scale datasets MidEast-TE and GDELT-TE. Finally, we proposed a novel forecasting method LoGo that outperforms SOTA methods by a large margin on both datasets. 

Several future directions warrant attention: improving the zero-shot event extraction by LLMs, exploring automatic schema induction from data, and developing advanced graph models for SCTc-TE forecasting are key areas for further research.

\bibliographystyle{ACM-Reference-Format}
\bibliography{0_main}

\end{document}